\documentclass[pre,aps,twocolumn]{revtex4-2}
\usepackage{ulem}
\usepackage[T1]{fontenc}
\usepackage{amsfonts}
\usepackage{times}
\usepackage{amsmath, amsthm}
\usepackage[final]{graphicx}
\usepackage{amssymb, algorithm, algpseudocode}
\algnewcommand\algorithmicforeach{\textbf{for each}}
\usepackage{xcolor}
\algdef{S}[FOR]{ForEach}[1]{\algorithmicforeach\ #1\ \algorithmicdo}
\usepackage[title]{appendix}
\definecolor{orange}{rgb}{1,0.5,0}
\usepackage{xcolor,comment}

\begin{document}
\title{Dispersive shock waves in periodic lattices}

\author{Su Yang}
\affiliation{Department of Mathematics and Statistics, University
of Massachusetts, Amherst, Massachusetts 01003-4515, USA}

\author{Sathyanarayanan Chandramouli}
\affiliation{Department of Mathematics and Statistics, University
of Massachusetts, Amherst, Massachusetts 01003-4515, USA}

\author{P. G. Kevrekidis}
\affiliation{Department of Mathematics and Statistics, University
of Massachusetts, Amherst, Massachusetts 01003-4515, USA}
\affiliation{Department of Physics, University of Massachusetts Amherst, Amherst, MA 01003, USA}
\affiliation{Theoretical Sciences Visiting Program, Okinawa Institute of Science and Technology Graduate University, Onna, 904-0495, Japan}
\date{\today}

\begin{abstract}
We introduce and systematically investigate the generation of dispersive shock waves, which arise naturally in physical settings such as optical waveguide arrays and superfluids confined within optical lattices. The underlying physically relevant model is a nonlinear Schrödinger (NLS) equation with a periodic potential. We consider the evolution of piecewise smooth initial data composed of two distinct nonlinear periodic eigenmodes. To begin interpreting the resulting wave dynamics, we employ the tight-binding approximation, reducing the continuous system to a discrete NLS (DNLS) model with piecewise constant initial data (i.e., a Riemann problem), where each constant state represents a discrete Floquet-Bloch mode at the continuum model level. 
The resulting tight-binding approximation is shown to display higher-fidelity for \textit{deeper} periodic potentials. This reduced DNLS model effectively models the dynamics at the minima of the periodic potential of the original continuum NLS. Within such a single-band DNLS framework, we apply tools from Whitham modulation theory and long-wave quasi-continuum reductions to uncover and analyze a rich spectrum of non-convex, discrete dispersive hydrodynamic phenomena, comparing the resulting phenomenology with that of the 
periodic-potential-bearing continuum model.
\end{abstract}

\maketitle

\section{Introduction}
Wave propagation in media with periodic heterogeneities has been of substantial interest in a broad range of disciplines. These include nonlinear optics (see, e.g., select experimental 
works in waveguide arrays \cite{szameit2006two,eisenberg1998discrete,iwanow2004observation}, and photonic crystal fibers \cite{argyros2005guidance,argyros2005photonic}, as well
as the broad review~\cite{Lederer2008_PhysRep_DiscreteSolitons}), Bose-Einstein condensates (BECs) trapped within an optical lattice (see e.g. the reviews \cite{brazhnyi2004theory,morsch}), fluid mechanics (e.g., describing shallow water propagation over a periodic bathymetry \cite{ketcheson2025multiscale,ketcheson2025dispersive}), nonlinear acoustics \cite{temple2023nonlinear,ketcheson2024solitary}), acousto-optics \cite{korpel1972acousto} and metamaterials (see, e.g., the books~\cite{Nester2001,yuli_book,granularBook}) besides many others. 

At its heart, wave propagation in such heterogeneous media is significantly different from the homogeneous case due to the dispersion introduced from the lattice (lattice diffraction/dispersion). The induced diffraction often exhibits remarkable properties, making it amenable to precise engineering in nonlinear optics \cite{eisenberg2000diffraction}. The interplay between nonlinearity and engineered lattice diffraction therein was shown to open various possibilities for, e.g., the existence of both bright and dark solitary waves \cite{peschel1998discrete,darmanyan1998strongly}. The diffraction management technique has also been successfully applied in the ultracold atomic gas setting, where a
significant degree of controllability arises from the Feshbach resonance technique~\cite{Chin2010Feshbach} which can modify the underlying nonlinear response as well \cite{brazhnyi2004theory}.
The manipulation of dispersion in such atomic BECs has also been responsible for the
creation of bright matter wave solitons in the setting of repulsive interactions
(so-called gap solitons) in the work of~\cite{markus}.
In a different vein, in the context of shallow water wave propagation, novel effects related to the induced lattice dispersion lead to the generation of traveling breathers \cite{ketcheson2025dispersive,ketcheson2025multiscale}. Similar phenomenology has also been reported in non-isentropic gas dynamics (e.g. \cite{ketcheson2024solitary}), wherein there is a counterintuitive formation of \textit{dispersive shock waves}.

The nonlinear Schrödinger (NLS) equation (alias Gross–Pitaevskii equation) with a periodic potential is a model of broad significance and wide applicability both in its continuum form~\cite{brazhnyi2004theory,Pelinovsky2011Localization}, as well as in its discrete analogue, i.e., the
so-called discrete NLS (DNLS)~\cite{kevrekidis2009discrete}. It 
can be thought of as a {\it generic envelope model}
describing the transverse "evolution" of weakly nonlinear, slowly varying optical wave envelopes in a waveguide array and serves as a direct approximation to the condensate order parameter in dilute ultracold atomic gases in the realm of the so-called mean-field
approximation in the presence of an optical lattice~\cite{PitaevskiiStringari2016}. Owing to its central role in both optics and ultracold atomic physics, it has been the focus of extensive investigations into the emergence of coherent structures. This body of work spans several decades, from early studies and reviews \cite{brazhnyi2004theory,morsch,Pelinovsky2011Localization} to very recent contributions in 
the experimental realm~\cite{cruickshank2025single}, which have provided
a fertile platform for visualizing coherent nonlinear waveforms in
such systems. Parallel developments in nonlinear optics have also been substantial 
and extend from earlier investigations of solitonic and vortical 
waveforms~\cite{kevrekidis2009discrete} to more recent ones on the interplay of
topology and discreteness~\cite{SzameitRechtsman2024DiscreteNonlinearTopologicalPhotonic}.

In a slightly different but related vein, a significant body of recent work in nonlinear dispersive media has focused on dispersive shock wave (DSW) generation ~\cite{hoefer_dispersive_2009,el_dispersive_2016-1}. These nonlinear wave patterns emerge due to competing dispersive and finite amplitude effects. DSWs have been observed in diverse physical settings, ranging from fluid flows~\cite{trillo_observation_2016,maiden_observation_2016,el_transformation_2012,ablowitz_nonlinear_2011,ablowitz_nonlinear_2012,fibich2015nonlinear} and optical media~\cite{wan_dispersive_2007,bendahmane_piston_2022,fatome_observation_2014}, to ultracold atomic gases~\cite{hoefer_dispersive_2006,chang_formation_2008,hoefer_matter-wave_2009,mossman2018dissipative} and engineered materials such as granular crystals~\cite{herbold,molinari_stationary_2009,granularBook} and Fermi–Pasta–Ulam–Tsingou-type magnetic lattices~\cite{li2021observation}. Within theoretical studies, the one-dimensional, homogeneous nonlinear Schrödinger equation and its variants have served as a primary \textit{bi-directional} framework for analyzing DSW formation and dynamics, motivating sustained research over several decades~\cite{el_dispersive_2016-1,el1995decay,hoefer_shock_2014,el_resolution_2005}. Following this line of development, the extension to \textit{discrete} Schrödinger-type systems became a natural next step \cite{kamchatnov2004dissipationless,mohapatra2025dam,panayotaros2016shelf,salerno2000shock,konotop1997dark}, driven by the numerous applications discussed above.

Although DSWs are universal features of nonlinear dispersive media, their investigation in periodic lattices is a topic that is relatively recent and has received somewhat 
limited attention. In this context, coupling effects between adjacent potential wells become particularly relevant. Experiments have demonstrated that Bloch mode coupling drives inter- and intraband energy transfer during DSW propagation \cite{jia2007dispersive}. 
Moreover, in engineered metamaterial lattices (monomer~\cite{herbold,HEC_DSW} and dimer~\cite{molinari_stationary_2009}, granular~\cite{granularBook} and
magnetic~\cite{li2021observation}), such shock waves have been observed to
spontaneously arise in a wide range of related experiments.
Despite the generality of these findings, valid across a range of coupling strengths between neighboring waveguides, a systematic  theory has yet to be established, although
a number of steps have been made recently in the direction of connecting 
such discrete problems with integrable~\cite{DSW2}, as well as quasi-continuum
descriptions amenable to modulation theory analysis~\cite{mohapatra2025dam,Yang2025RegularizedContinuum}.

In this work, we revisit this problem of shock wave generation by examining the dynamics of “generalized” Riemann problems (see \cite{glimm1984generalized} for a definition of this notion), defined by piecewise-smooth initial data consisting of two distinct nonlinear periodic lattice eigenmodes. Our framework relies on expanding the wavefunction in a \textit{complete} basis of localized Wannier functions and employing the tight-binding approximation \cite{Alfimov_2002}. 
This type of approach has been previously successfully used for ``translating'' the
continuum problem with periodic potential to the DNLS one for the case of solitary waves~\cite{Alfimov_2002}. However, 
it has not been deployed as an approach for understanding dispersive shock waves in the
experimentally tractable, spatially periodic NLS with an optical lattice, to the best of
our knowledge.
By controllably neglecting long-range interactions between distant waveguides, this approximation provides a first step toward uncovering the mechanisms governing DSW generation and propagation. The resulting model can then be analyzed using recent developments in non-convex dispersive hydrodynamics \cite{kamchatnov2004dissipationless,sprenger2023whithammodulationtheorytwophase,mohapatra2025dam}, offering insights into the observed shock features. {Non-convex (or non-classical) dispersive hydrodynamics is an emerging field in which wavetrains distinct from ``classical", KdV-type dispersive shock waves \cite{gurevich1974nonstationary} are generated due to either
\begin{enumerate}
    \item The non-convexity of the dispersionless limit: for scalar PDEs this corresponds to a non-convex (nonlinear) flux, while for dispersive hydrodynamic systems this generalizes to either the loss of  the so-called ``strict" hyperbolicity" or the loss of genuine nonlinearity of the dispersionless limit (c.f. \cite{van2021dynamic,el2017dispersive,ivanov2020riemann,kamchatnov2012undular}). Both notions are discussed also in Sec.~\ref{tightb}. From a phenomenological standpoint, the loss of genuine nonlinearity is associated with the preferential formation of contact discontinuities rather than classical shock waves in the associated, dispersionless limit. Moreover, it can preclude the existence of simple-wave solutions, such as rarefaction waves.
    \item The non-convexity of linear dispersion: here, there exist specific wavenumbers when the dispersion sign ${\rm sgn}[{\rm det}(\partial_{k_i k_j}\omega)]$ is zero, where $\partial_{k_i k_j}\omega$ is the Hessian associated with the linear dispersion relation $\omega(\Vec{k})$. Representative works discussing such dynamics include, but are not limited to \cite{sprenger2017shock,baqer2025shallow,congy2021dispersive}.
\end{enumerate}
DSWs in lattice systems typically possess a non-convex and band-limited linear dispersion relation \cite{sprenger2024hydrodynamics,turner1997small,mohapatra2025dam}, which underpins the rich array of non-classical phenomena observed in such settings.
}

We also identify regimes—accessible through controlled parameter variation—in which the 
single-band tight-binding approximation breaks down, primarily because it fails to capture energy transfer between admissible bands. These considerations pave the way toward a deeper and more comprehensive understanding of shock generation in continuum dispersive models bearing periodic lattice potentials.

Our work is organized as follows. In Sec.~\ref{Section-1}, we introduce the underlying model and the initial conditions used to study dispersive shock wave (DSW) generation. We then discuss, in Sec.~\ref{Section1C}, the derivation of the class of tight-binding approximations and their corresponding initial data. Section~\ref{tightb}
describes the phenomenology at the lattice level, while section~\ref{wavepatt} discusses
the implications of the tight binding findings towards the phenomenology of the
continuum model with a periodic potential. Finally, section~\ref{conclusions} summarizes
our findings and presents some interesting directions for future study.

\section{Model description and initial condition}
\label{Section-1}
\subsection{Model description}
We consider the nonlinear Schrödinger equation (NLS), which in the context of superfluids is commonly referred to as the Gross–Pitaevskii equation (GP)~\cite{brazhnyi2004theory,morsch,PitaevskiiStringari2016}:
\begin{equation}\label{eq: Periodic dNLS}
    i\psi_t = -\psi_{xx} + V\left(x\right)\psi +\left|\psi\right|^{2}\psi,
\end{equation}
where $V(x)$ is a $L$-periodic potential function ($V\left(x+L\right) = V\left(x\right)$). 
This model admits two conserved quantities: the mass (or particle number),
\begin{equation}
    N[\psi]=\int |\psi(x,t)|^2 dx,
\end{equation}
and the ``energy" (or Hamiltonian)
\begin{equation}
    E[\psi]=\int \frac{1}{2} \left(|\psi_x|^2+V(x)|\psi|^2+ |\psi|^4\right) dx,
\end{equation}
{where an additional potential energy term arises due to the periodic potential, representing the extra energy barrier that waves must overcome.}

In the context of spatial optics (ultracold Bose gases), the periodic potential can be experimentally realized using a lattice of waveguides engineered via a pair of ordinarily polarized plane waves \cite{jia2007dispersive} (laser-induced optical lattices \cite{anderson1998macroscopic}). Another significant application of Eq.~\eqref{eq: Periodic dNLS} is found in photonic crystals, where the same mathematical framework describes the propagation of light in nonlinear, periodic dielectric structures \cite{soukoulis2012photonic}.
To illustrate specific features of the system, we focus on a prototypical periodic potential given by
\begin{equation}\label{eq: Periodic potential}
    V\left(x\right) = V_0\sin^{2}\left(x\right),
\end{equation}
where $V_0\in\mathbb{R}$ is a real constant.  For $|x|\ll 1$, the potential behaves as $V(x)\sim V_0 x^2$, effectively mimicking a harmonic trap, which is common in Bose–Einstein condensate (BEC) experiments and their theoretical modeling \cite{pethick2008bose,kevrekidis2008emergent,sharan2025breaking}. 

The NLS equation admits an underlying dispersive-hydrodynamic interpretation, which can be seen by adopting the polar representation (the so-called Madelung transformation) for the wavefunction {$\psi=\sqrt{\rho(x,t)}\exp(i\Phi(x,t))$}, {where $u=\Phi_x$,}
\begin{align}
\label{eqn5}
   & \rho_t+(2\rho u)_x=0\;\\\nonumber
   & u_t+2uu_x+\rho_x=-V_x+\left(\frac{2\rho \rho_{xx}-\rho_x^2}{4\rho^2}\right)_x.
\end{align}
Notably, the left-hand side of the system of equations resembles the shallow-water equations, {where $\Phi(x,t)$ represents the ``fluid" velocity potential, while its gradient $u=\partial_x\Phi$ represents the associated hydrodynamic velocity. Furthermore, this shallow water system is known to develop gradient singularities in finite time.} 
Here, the combined effects of media heterogeneity and the dispersive quantum pressure terms
---the two terms on the right hand side of the 2nd one of Eqs.~(\ref{eqn5})---
must act in order to avert the unphysical gradient catastrophe in the lattice system. 
It is important to recall that even just the quantum pressure term is sufficient to that 
effect~\cite{PitaevskiiStringari2016}.

In the absence of the potential $V(x)\equiv 0$, modulated periodic wavetrains known as dispersive shock waves (DSW) are formed. DSWs are characterized by two length scales: the coherence scale \cite{el_dispersive_2016-1} (typical scale of nonlinear periodic oscillations) and the hydrodynamic scale (modulation/slow variation across several nonlinear periodic oscillations). These rapidly oscillatory, modulated structures connect two distinct homogeneous backgrounds supported by the bulk medium.

However, comparatively few quantitative studies have explored dispersive hydrodynamics under the combined influence of quantum pressure and periodic heterogeneity {\it on DSWs} (cf. the experimental investigation in \cite{jia2007dispersive}), a gap that the present work seeks to ---at least partially--- address. In particular, we extend the concept of dispersive shock waves (DSWs) 
to continuum but periodic systems, due to the connection of the latter with the
Riemann problem on a discrete lattice. A key framework enabling this generalization is the tight-binding approximation, which we will now detail.

\subsection{Band structure and Initial condition}
\begin{figure}
    \centering
    \includegraphics[width=\linewidth]{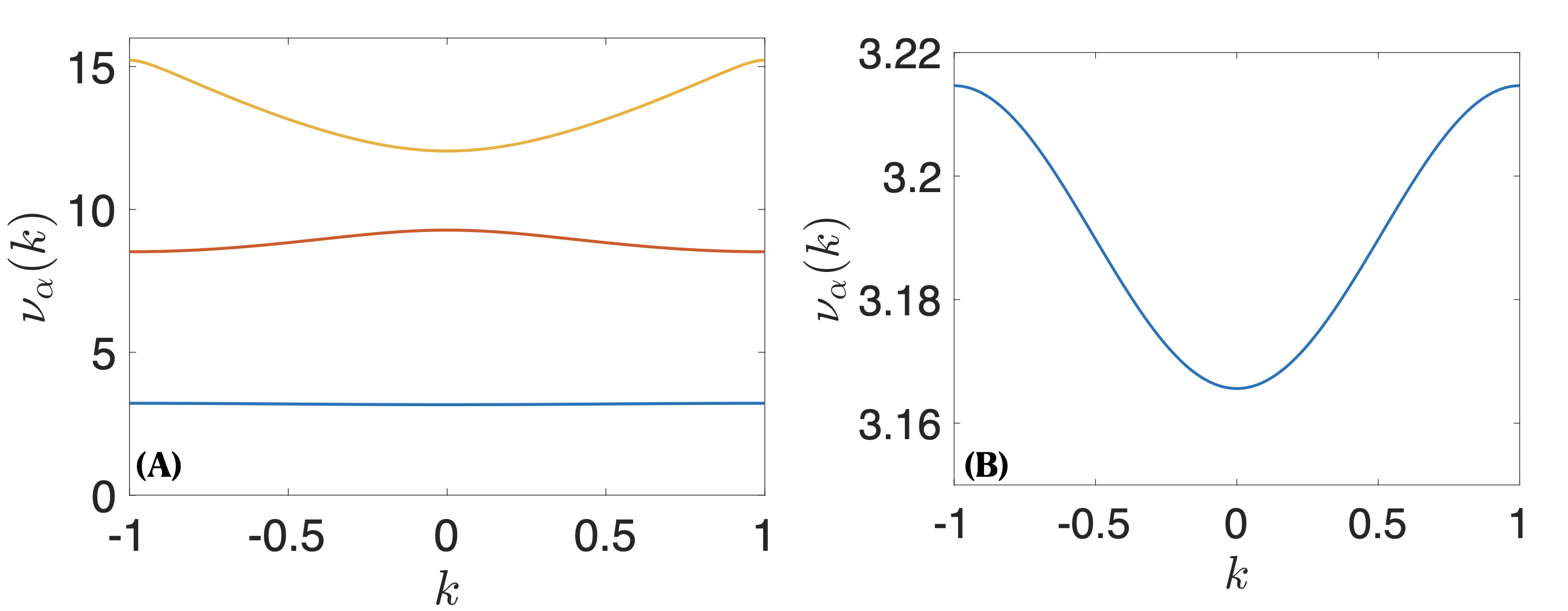}
    \caption{(a) The band structure for the linear Schrödinger equation with potential $12\sin^2(x)$, displaying the first three bands shown in the first Brillouin zone, (b) Zoom on the lowest eigenvalue band showing the change in curvature (diffraction sign) between the band edge at $k=0$ and the edge points $k=\pm1$. }
    \label{fig:bs}
\end{figure}

At its core, wave propagation in a $L$-periodic, heterogeneous medium is fundamentally altered by the dispersion introduced by the lattice (lattice diffraction), which creates allowed energy bands and forbidden gaps in the linear spectrum. To fix ideas, without loss of generality, we consider the linear Schrödinger operator defined by 
\begin{equation}
    -\frac{d^2\phi_{k,\alpha}}{dx^2}+V(x)\phi_{k,\alpha}(x)=\nu_{\alpha}(k)\phi_{k,\alpha}(x),
    \label{Lin-diffraction-relation}
\end{equation}
where $\phi_{k,\alpha}(x)$ has $L$-periodic Bloch-Floquet functions $\phi(x)=\varphi(x)\exp(ikx)$, $k\in[-\pi/L,\pi/L]$ (Bloch wavenumber) and the eigenvalue (energy) bands $\nu_{\alpha}(k)$ are $2\pi/L$ periodic in the Bloch wavenumber, i.e., can be written as a Fourier series
\begin{equation}
    \nu_{\alpha}(k)=\sum_n \hat{\omega}_{n,\alpha} e^{iknL},
\end{equation}
where we have the symmetry $\hat{\omega}_{n,\alpha}=\hat{\omega}_{-n,\alpha}=\hat{\omega}^*_{n,\alpha}$.
In Fig.~\ref{fig:bs}, we present the first three eigenvalue bands—separated by band gaps that correspond to forbidden propagation zones—for the potential $V(x)=V_0\sin^2(x)$ with $V_0=12$, using the reduced Brillouin zone representation.
Furthermore, the dispersion or diffraction curvature at $k=0$ ($k=\pm1$) has a positive (negative) sign. When coupled to cubic nonlinearity in Eq.~\eqref{eq: Periodic dNLS}, this points to effectively defocusing (focusing) wave dynamics. In particular, we are interested in exploring the regimes associated with such defocusing media.
\begin{figure}
    \centering
    \includegraphics[width=\linewidth]{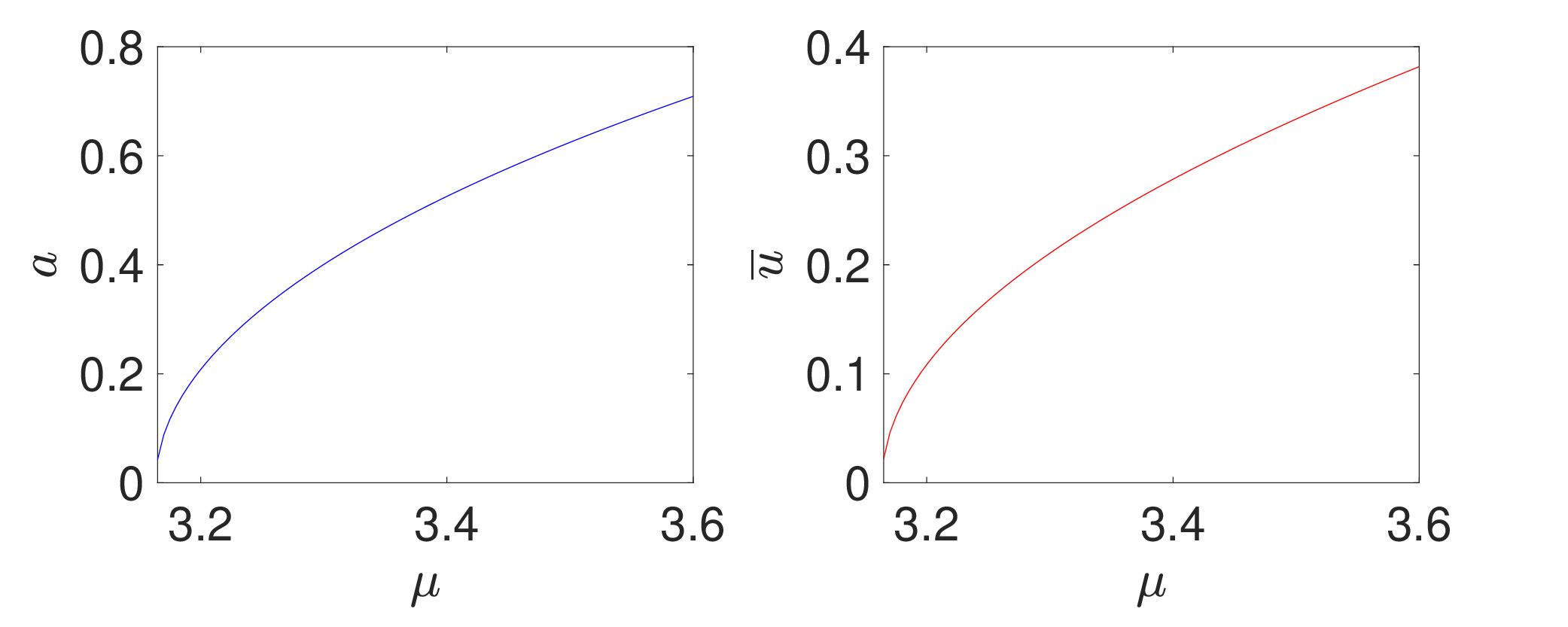}
    \caption{The (left) amplitude and (right) wavemean variations of the nonlinear eigenmodes of Eq.~\eqref{eq: Nonlinear steady-state problem} as a function of the eigenvalue $\mu$, for $V_0=12$.}
    \label{fig:Amp_wave_mean}
\end{figure}
Thus, bifurcating from the $k=0$ edge, we seek steady, nonlinear eigenmodes to Eq.~\eqref{eq: Periodic dNLS} by assuming the ansatz
\begin{equation}\label{eq: steady state ansatz}
\psi\left(x,t\right) = u\left(x\right)\exp\left(-i\mu t\right),
\end{equation}
where $\mu \in \mathbb{R}$ denotes the value of the chemical potential (for atomic condensates)/ propagation constant (for optical waveguide arrays) and $u(x)$ is a real valued function. The steady states satisfy the nonlinear eigenvalue problem given below
\begin{equation}\label{eq: Nonlinear steady-state problem}
    -u_{xx}+V\left(x\right)u+u^{3} = \mu u.
\end{equation}

Clearly, Eq.~\eqref{eq: Nonlinear steady-state problem} does not possess homogeneous states (plane waves) as solutions. Instead, there exists a one-parameter family of nonlinear periodic states, bifurcating from the zero-momentum limit of the \textit{first band}.
Explicit formulas are not available for such eigenmodes
[except in the cases of special cnoidal wave potentials~\cite{carr1,carr2}], but they can be computed using a standard Newton's method (given a value of $\mu$) with periodic boundary conditions enforced. 
The amplitude and wavemean variations within this family are shown as a function of $\mu$ in Fig.~\ref{fig:Amp_wave_mean}, for $V_0=12$ (W.L.O.G). 


In analogy with the classical Riemann problem used to study the evolution of disturbances on homogeneous backgrounds in the bulk NLS~\cite{el1995decay}, we introduce a generalized Riemann problem to describe the evolving modulations of nonlinear periodic states. The underlying model \eqref{eq: Periodic dNLS} satisfies a reflection symmetry ($V(x)$ considered here possesses an even parity), while possessing only a \textit{discrete} (shift) translational 
invariance. The underlying symmetries (or lack thereof) suggest a three-parameter family of Riemann problems, consisting of two distinct periodic states and the location of the transition between the states (relative to the minima of the periodic potential).

However, in this work, we define a two-parameter family of Riemann problems with two distinct values of $\mu$ to obtain a pair of steady states denoted by $u^{-}\left(x\right)$, $u^{+}\left(x\right)$, respectively. Thus, we construct the \textit{generalized} Riemann data (c.f. \cite{glimm1984generalized}) for Eq.~\eqref{eq: Periodic dNLS} as follows,
\begin{equation}\label{eq: Initial condition for inhomogeneous NLS}
   \psi\left(x,0\right) = \begin{cases}
       u^{-}\left(x\right), \hspace{5mm} x <n_0L,\\
       u^{+}\left(x\right), \hspace{5mm} x> n_0L,
   \end{cases}
\end{equation}
where {$n_0$} indicates the $n_0^{th}$ lattice site. We further restrict our investigation by fixing $\mu_{-}>\mu_{+}$ and holding $\mu_+$ as a constant. Furthermore for computational tractability, the transition between the nonlinear eigenmodes is smoothened, and is specified to occur across 3 potential minima.
We perform the time stepping Eq.~\eqref{eq: Periodic dNLS} with the initial data defined in \eqref{eq: Initial condition for inhomogeneous NLS} by a pseudospectral ETDRK4 \cite{kassam2005fourth} method.
\subsection{Wannier basis reduction and the tight-binding reduction}
\label{Section1C}
To interpret the dynamics of the generalized Riemann problems, we employ the well-known, (complete and orthonormal) Wannier basis representation of $\psi(x,t)$, which has been extensively discussed in the {context of lattice waveguide systems and also the propagation of nonlinear waves therein (although not in that of
DSWs, to our knowledge)~\cite{Alfimov_2002,cole2024data,kevrekidis2009discrete,brazhnyi2004theory}}:
\begin{equation}\label{eq: Compleness of WFs}
    \psi\left(x,t\right) = \sum_{n}\sum_{\alpha}c_{n,\alpha}\left(t\right)w_{n,\alpha}\left(x\right),
\end{equation}
where $n$ denotes the lattice site about which each of these basis functions $w_{n,\alpha}(x)$ is localized, $\alpha$ represents the band index, and $c_{n,\alpha}(t)$ are the time-dependent coefficients in the expansion.

Substituting the expansion into Eq.~\eqref{eq: Periodic dNLS}, we obtain the \textit{exact}
---although, at first glance, considerably more complex--- system \begin{align}
\label{system-wannercomplex}
    &i\frac{dc_{n,\alpha}}{dt} = \sum_{n_1}c_{n_1,\alpha}\hat{\omega}_{n-n_1,\alpha} + \\\nonumber &\sum_{\alpha_1,\alpha_2,\alpha_3}\sum_{n_1,n_2,n_3}c_{n_1,\alpha_1}^{*}c_{n_2,\alpha_2}c_{n_3,\alpha_3}w_{\alpha\alpha_1\alpha_2\alpha_3}^{nn_1n_2n_3},
\end{align}
{where the contributions from the linear terms in Eq.~\eqref{eq: Periodic dNLS} yield the discrete convolution in the system Eq.~\eqref{system-wanner}, while} the so-called the overlap matrix elements are defined by
\begin{align}\label{eq: Overlapping matrix elements}
&w_{\alpha\alpha_1\alpha_2\alpha_3}^{nn_1n_2n_3} \\\nonumber&= \int_{\mathbb{R}}w_{n,\alpha}\left(x\right)w_{n_1,\alpha_1}\left(x\right)w_{n_2,\alpha_2}\left(x\right)w_{n_3,\alpha_3}\left(x\right)dx.
\end{align}
However, this framework of coupled discrete equations provides a means to begin analyzing the dynamics of the underlying heterogeneous PDE of Eq.~\eqref{eq: Periodic dNLS}, subject to the initial condition in Eq.~\eqref{eq: Initial condition for inhomogeneous NLS}. As explicated in \cite{Alfimov_2002}, the coupled discrete system governing the evolution of the Wannier basis coefficients can be simplified by assuming that the Wannier functions $w_{n,\alpha}(x)$ are strongly localized, which yields a reduced form of the overlap matrix elements, approximated as $\approx w^{nnnn}_{\alpha_1\alpha_2\alpha_3\alpha_4}$, upon ignoring contributions from lattice sites at $n_1$, $n_2$ and $n_3$, as being exponentially weaker when these indices
are not identical to $n$. This, 
in turn, modifies {Eq.~\eqref{system-wannercomplex} to}:
\begin{align}
\label{system-wanner}
    &i\frac{dc_{n,\alpha}}{dt} = \sum_{n_1}c_{n_1,\alpha}\hat{\omega}_{n-n_1,\alpha} + \\\nonumber &\sum_{\alpha_1,\alpha_2,\alpha_3}c_{n,\alpha_1}^{*}c_{n,\alpha_2}c_{n,\alpha_3}w_{\alpha\alpha_1\alpha_2\alpha_3}^{nnnn}.
\end{align}
A further approximation can be made by assuming only intraband energy transfer (only within $\alpha=1$), resulting in
\begin{align}
\label{system-wanner-full}
    &i\frac{dc_{n,1}}{dt} = \sum_{n_1}c_{n_1,1}\hat{\omega}_{n-n_1,1} +w_{1111}^{nnnn}|c_{n,1}|^2c_{n,1}.
\end{align}
This approximation implies that the potential well is such (e.g., sufficiently deep) that
the first band is energetically quite distant from higher bands, and thus a single-band
description adequately captures the dynamics. Notice that this can be suitably
modified to include the cross-talking of more bands in a vector version of the
{discrete nonlinear Schrödinger} equation~\cite{Alfimov_2002}, if this is dictated by the energetics of the relevant
periodic potential.

\begin{figure}
    \centering
    \includegraphics[width=\linewidth]{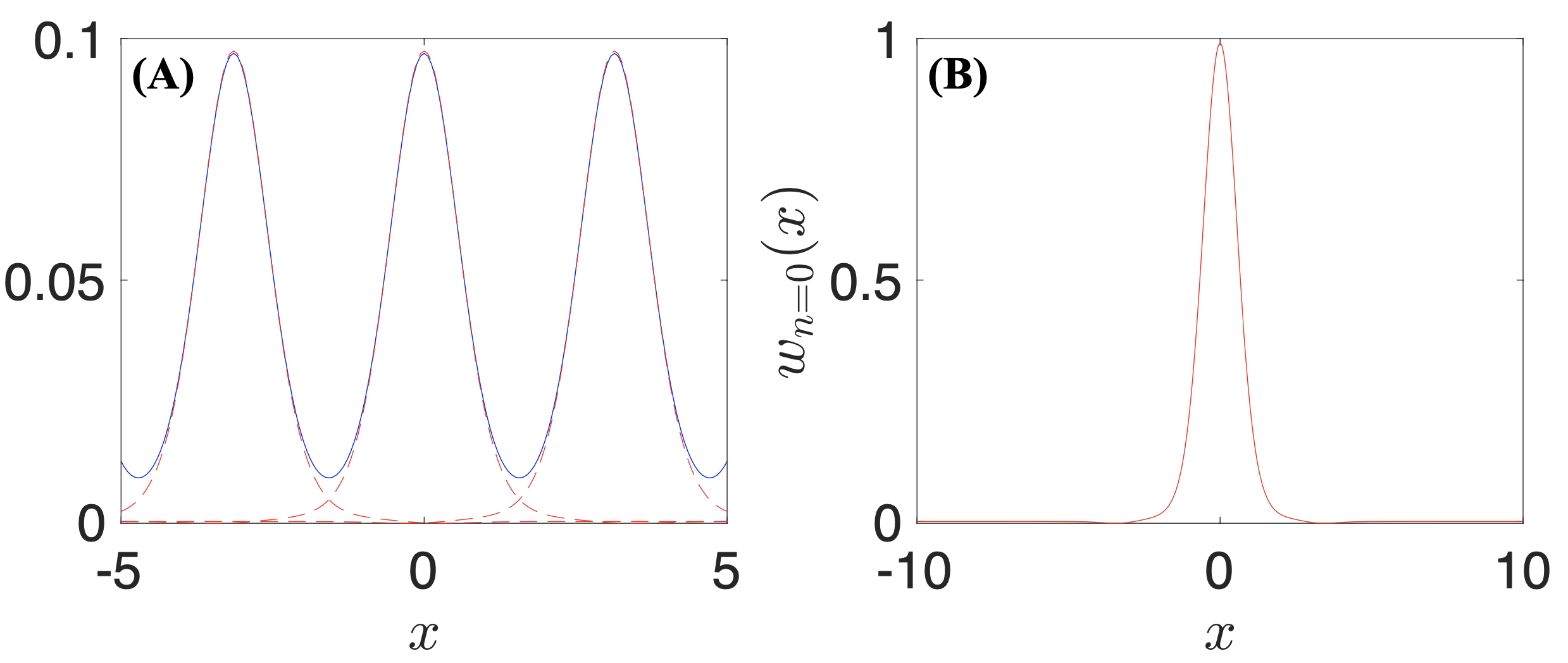}
    \caption{The periodic eigenmode ($u(x)$) with eigenvalue $\mu=3.17$ constructed by superposing first-band Wannier functions for $V_0=12$. The periodic state is shown with blue solid color, while the scaled and shifted Wannier basis functions are shown in red-dashed line (A). (B) A single Wannier basis function centered at $n=0$. For $V_0=12\gg1$, the basis function is nearly Gaussian-like \cite{Alfimov_2002}.}
    \label{fig:periodic_construction}
\end{figure}

Furthermore, for $V_0\gg1$,  the rapid decay of the Fourier coefficients $\hat{\omega}_{n,1}$ can be verified \cite{kevrekidis2009discrete,Alfimov_2002}, leading to a simplification of the discrete convolution term to include contributions only from the nearest lattice sites ${n\pm 1}$ to yield
\begin{align}
\label{system-wanner}
    &i\frac{dc_{n,1}}{dt} = \hat{\omega}_{0,1} c_{n,1} + \hat{\omega}_{1,1}(c_{n-1,1}+c_{n+1,1})\\\nonumber&+w_{1111}^{nnnn}|c_{n,1}|^2c_{n,1},
\end{align}
also referred to as the first- or lowest-band  tight-binding approximation. {We refer to Eq.~\eqref{system-wanner} as the scalar DNLS in our present work.} In this regime, the propensity for tunneling across adjacent lattice sites is minimal (since $\hat{\omega}_{1,1}\ll\hat{\omega}_{0,1}$). Furthermore, the initial condition for the continuum problem in this case reduces to a \textit{dam-break} problem
\begin{equation}
\label{eq:riemann}
    c_n(0)=\begin{cases}
        c^{-}, \;n<n_0\\
        c^{+},\; n>n_0,
    \end{cases}
\end{equation}
{where $c^{\pm}$ are} the corresponding right (and left) \textit{discrete} Floquet-Bloch modes and $c^{-}>c^{+}$, since we restrict attention to $\mu_->\mu_+$ in Eq.~\eqref{eq: Initial condition for inhomogeneous NLS}.
Notice that a key point that we leverage herein is that the seemingly far more
complex dam break problem involving two Floquet-Bloch modes
in the heterogeneous, continuum PDE simplifies considerably to a ``regular'' 
Riemann problem at the DNLS level. Hence, our plan is to address the latter,
utilizing ideas from~\cite{mohapatra2025dam}, among others, in order to obtain
information about the continuum problem with the periodic potential.
\\\textbf{Remark:} Restricting consideration to the first band, we must have 
\begin{equation}
\label{eq:superposn-elementary}
\psi^{\mp}(x,0)\approx c^{\mp}\sum_{n}w_{n}(x),\end{equation} and thus, $c^{\mp}\approx \psi^{\mp}(x,0)/(\sum_{n}w_{n}(x))$. 
Fig.~\ref{fig:periodic_construction} illustrates this trivial superposition of first band Wannier basis functions to yield the lattice eigenmodes.


\section{Dispersive hydrodynamics for the tight-binding approximation}
\label{tightb}
\subsection{The one-phase modulation equations and simple waves}
The interpolating model for the first tight-binding approximation is a (band-limited) full-dispersion NLS \cite{sprenger2023whithammodulationtheorytwophase}
\begin{equation}
\label{Interpolating-DNLS}
    i\Psi_t=\Tilde{\Omega}(-i\partial_x)\Psi+w_{11}|\Psi|^2\Psi
\end{equation}
where $\Psi(nL,t)=c_n(t)$ and pseudodifferential operator is defined in Fourier space as
\begin{equation}\label{eq: Full dispersion relation}
\mathcal{F}[\Tilde{\Omega}]=\Omega\left(k;L\right) = 2\hat{\omega}_{1}\cos\left(k L\right) + \hat{\omega}_{0}.
\end{equation}
\textbf{Remark:} The band-index in the Fourier coefficients $\hat{\omega}_j$ has been dropped with the implicit understanding that we are restricting attention to the first band alone. For the same reason, we have also adopted a simplified notation for the nonlinear coefficient $w_{11}$ (the overlapping matrix element) \cite{Alfimov_2002}. For the potential $V=V_0 {\rm sin}^2(x)$ in consideration, ${\rm sgn}(\hat{\omega}_0)={\rm sgn}(w_{11})=-{\rm sgn}(\hat{\omega}_1)=1$.

Notably, $\Omega$ provides a good approximation to the linear dispersion relation of the first band in Eq.~\eqref{eq: Periodic dNLS}, i.e., $\nu_{1}(k)$, provided that the coefficients of the higher modes decay sufficiently rapidly ($|\hat{\omega}_{j+1}|\ll |\hat{\omega}_{j}|$ for all $j$). This in particular is expected for sufficiently large $V_0\gg 1$ (see for eg. \cite{Alfimov_2002}). 

Stokes wave solutions (one-phase periodic solutions) to Eq.~\eqref{Interpolating-DNLS} can be written down explicitly as
\begin{equation}
    \label{Stokes-waves}
    \Psi_0(x,t)=\sqrt{\rho_0}\exp\Bigg(i\bigg(k x-\left(\Omega(k)+w_{11}\rho_0\right)t\bigg)\Bigg).
\end{equation}

Slow modulations of Stokes waves (with density $\rho(X,T)$ and wavenumber $\overline{k}(X,T)$) to Eq.~\eqref{Interpolating-DNLS} can be cast in Riemann invariant form \cite{sprenger2023whithammodulationtheorytwophase} to yield
\begin{equation}
    \label{One-phase-WM-system}r^{(1,2)}_T+\lambda^{(1,2)}r^{(1,2)}_X=0,
\end{equation}
 where $T\sim \epsilon t$, $X\sim \epsilon x$ and $\epsilon \ll 1$. The expressions for these Riemann invariants and associated characteristic velocities are given by
 \begin{equation}
    r^{(1,2)} = 2\sqrt{2}\sqrt{-\hat{\omega}_1}E\left(\frac{\overline{k} L}{2},2\right) \pm 2\sqrt{w_{11}}\sqrt{\overline{\rho}},
\end{equation}
where {$E(\varphi,m=2)=\int_{0}^{\varphi} \sqrt{1-2\sin^2(\theta)} d\theta$ }represents the {incomplete elliptic integral of the second kind evaluated at the elliptic modulus parameter $m=2$,} and \begin{equation}
    \lambda^{(1,2)} = -2L\hat{\omega}_1\sin(\overline{k} L) \pm \sqrt{-2\hat{\omega}_1w_{11}L^2\overline{\rho}\cos\left(\overline{k} L\right)}
\end{equation}
 respectively. 
 Recall that here $L$ is the periodicity scale of the external potential, which amounts
 to the effective spacing within our lattice considerations.
 Here, $\overline{k}$ and $\overline{\rho}$ are the slowly varying wavenumber and density respectively. Important properties associated with this modulation system include (i) the \textit{strict hyperbolicity} and (ii) its \textit{genuine nonlinearity}. Strict hyperbolicity is associated with the existence of real-valued, and strictly ordered characteristic velocities $\lambda_1<\lambda_2$, which is ensured when $\overline{k}L<\pi/2$. On the other hand, the genuine nonlinearity is ensured when $\nabla_{[\overline{\rho},\overline{k}]} \lambda^{(1,2)}\cdot {\bf R}^{(1,2)}(\overline{\rho},\overline{k})\neq 0$, where ${\bf R}^{(1,2)}$ constitute the pair of right-eigenvectors associated with the characteristic speeds \cite{dafermos1999genuinely}. 
{The existence of simple waves ({e.g.} rarefaction waves) to this modulation system depends on both its (i) strict hyperbolicity and (ii) genuine nonlinearity.} In particular, as was shown in our previous work \cite{mohapatra2025dam}, the breakdown of hyperbolicity is the primary determining factor for the non-existence of rarefaction waves. To explicate this, we consider the possible left (or slow)-propagating rarefaction waves, i.e., 1-wave, that could resolve the Riemann initial data in Eq.~\eqref{eq:riemann}. In particular, this wave is the long-time asymptotic solution to a Riemann initial condition with left Stokes wave $(c_{-},0)$ and right asymptotic Stokes wave $(\sqrt{\rho_0},k_0)$. 
{The 1-wave, or first rarefaction-wave family, is obtained  by holding the second Riemann invariant constant across its profile in the diagonal one-phase modulation system of Eqs.~\eqref{One-phase-WM-system}~\cite{el_dispersive_2016-1,el2007theory}. With this in mind, the variations in the first Riemann invariant are governed by the scalar, hyperbolic PDE $r^{(1)}_T+\lambda^{(1)}r^{(1)}_X=0$, while the (constant) second Riemann invariant satisfies $r_2(k_0,\rho_0)=r_2(0,(c^{-})^2)$. This latter relationship in turn defines a jump condition between the end states, }
thereby determining the relationship between them
\begin{equation}
\label{Intermediate-wave-dens}
    E\left(\frac{{k}_0L}{2},2\right)=\sqrt{\frac{-w_{11}}{2\hat{\omega}_1}}(|c^{-}|-\sqrt{{\rho_0}}).
\end{equation}
Given the monotonicity of the left-hand side (as a function of $k_0$), and that its domain (range) is restricted to $k\in (-\pi/2,\pi/2)$ ($\in (-0.59,0.59)$), the corresponding limited range (required for strict hyperbolicity) constrains the allowable jump $(|c^{-}|-\sqrt{\rho_0})$ that can give rise to rarefaction wave generation. The upper bound for the critical jump in $|c^{-}-\sqrt{\rho_0}|$ can be estimated to be $\sqrt{-\frac{2\hat{\omega}_1}{w_{11}}}E\left(\pi/4,2\right)$. {Evidently, as $\hat{\omega}_{1}$ decreases (corresponding to weaker coupling between neighboring waveguides), the critical jump also diminishes. }

\begin{figure}
    \centering
    \includegraphics[width=0.75\linewidth]{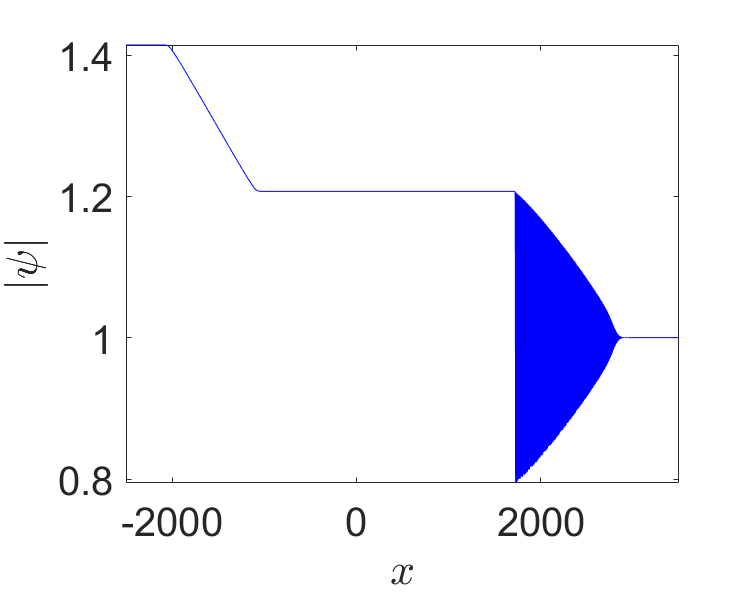}
    \caption{{The wavepattern at $t=1000$ arising from a Riemann problem ($\psi(x,0)=\sqrt{2},\;x<0$ and $\psi(x,0)=1,\;x>0$) posed to the continuum, bulk NLS (Eq.~\eqref{eq: Periodic dNLS} with $V(x)\equiv 0$). Here, a counter-propagating rarefaction wave and a DSW structure develop across a continuously expanding plane-wave background, facilitating the equilibration of hydrodynamic pressure across the initial step in wave amplitude $|\psi|$.}}
    \label{fig:bulk-NLS}
\end{figure}

{Riemann problems for the continuum bulk NLS ($V(x)\equiv 0$ in Eq.~\eqref{eq: Periodic dNLS}) are a canonical theme in the literature of convex, bi-directional dispersive hydrodynamics \cite{el2007theory,el1995decay,el_dispersive_2016-1}. In particular, from Riemann problems (only in the initial amplitude $|\psi(x,0)|$), an imbalance in the hydrodynamic pressure $\rho^2/2$ results, with $\rho=|\psi|^2$ being the hydrodynamic density. A counter-propagating rarefaction wave and a DSW pair develop across an expanding uniform-wave background (characterized by amplitude $\sqrt{\rho_0}$ and hydrodynamic velocity $k_0$), facilitating pressure equilibration (see Fig.~\ref{fig:bulk-NLS}). The DSW in this configuration emerges as the long-time asymptotic solution with left boundary data ($\sqrt{\rho_0},k_0$) and right boundary data ($|c^+|,0$) in the far-field. {In particular, it regularizes the multivalued structure generated in the dispersionless limit, while maintaining a constant first Riemann invariant ($r^{(1)}$) across its profile \cite{el2005undular} (referred to as the 2-DSW \textit{jump} condition).} Accordingly, since only the second Riemann invariant varies across its profile, this structure is termed a 2-DSW.}

{Next, we seek the parametric intervals in which a continuum, bulk NLS like wavepattern is obtained for the discrete tight binding model. To do so, we assume a 1-rarefaction wave (satisfying $r^{(2)}(k_0,\rho_0)=r^{(2)}(0,(c^{-})^2)$) and 2-DSW (satisfying $r^{(1)}(k_0,\rho_0)=r^{(1)}(0,(c^{+})^2)$) with a continuously expanding intermediate plane wave between them, characterized by amplitude $\sqrt{\rho_0}$ and one-phase wavenumber $k_0$. Matching these two constraints leads to the relationship 
\begin{equation}
    \label{density}\sqrt{\rho_0}=\frac{|c^-|+|c^+|}{2},
\end{equation}
and a jump condition for $|c^-|-|c^+|$, i.e.
\begin{equation}
    \label{velocity}
    |c^-|-|c^+|=2\sqrt{-\frac{2\hat{\omega}_1}{w_{11}}}E\left(\frac{k_0L}{2},2\right).
\end{equation}

Setting the argument of the monotone incomplete elliptic integral to $k_0=\pi/2$, yields an upper bound for this jump condition, yielding }
\begin{equation}
    |c^{-}|-|c^{+}|<2\sqrt{-\frac{2\hat{\omega}_1}{w_{11}}}E(\pi/4,2).
\end{equation}
 Notably, this upper bound for the jump leading to the existence of counterpropagating rarefaction and DSW diminishes for increasing $V_0$, as $-\hat{\omega}_1/w_{11}$ diminishes. Thus, in general, we expect a far richer landscape of dispersive
 hydrodynamic features ({besides the counterpropagating 1-rarefaction, 2-DSW pair}) to emanate from a Riemann problem, as demonstrated in our earlier work \cite{mohapatra2025dam}.

\subsection{{Linear dispersion relation of two-phase waves}}
{As a natural step to modulations of Stokes waves, we consider finite amplitude disturbances propagating on them, i.e., $c_n=\left(\sqrt{\rho_0}+a_n(t)\right)\exp\left(iknL-i\Omega t-w_{11}\rho_0 t-i\phi_n(t)\right)$. Applying an infinitesimal amplitude assumption on these distrubances, i.e. $a_n(t)=a_0 \exp(i(\kappa nL-\omega_0 t))$ ($a_0\ll 1$) and $\phi=\phi_0 \exp(i(\kappa nL-\omega_0 t))$ ($\phi_0\ll 1$), we obtain two branches of the dispersion relation associated with a \textit{second} phase of periodic oscillations (in line also with the analysis of~\cite{peyrard}})

 \begin{align}\label{eq: linear dispersion relation}
    &\omega_0^{\pm}(\kappa;k,\rho_0) = -2\hat{\omega}_1\sin(kL)\sin(\kappa L)\\\nonumber& \pm 2\sqrt{-\hat{\omega}_1}\sqrt{\cos(k L )}\sin\left(\frac{\kappa L}{2}\right)\times\\\nonumber &\sqrt{-4\hat{\omega}_1\cos(kL)\sin^{2}\left(\frac{\kappa L}{2}\right)+2{\rho}_0w_{11}},
\end{align}
where $\kappa L\in [-\pi,\pi] $ and $\omega_0$ are the associated wavenumber and angular frequency. The long wave expansion reveals parametric intervals (in $k$-$\rho$) in which competing higher-order dispersive (beyond the third-order) terms could become significant \cite{kamchatnov2004dissipationless,mohapatra2025dam}. Furthermore, there could exist inflection points in the dispersion, i.e., where $\partial_{\kappa\kappa}\omega_0^{\pm}=0$ \cite{mohapatra2025dam}. We bear this in mind, as this will have implications for propagating small amplitude shock waves.
\subsection{DSW Fitting}
We attempt to provide a global, quasi-analytical description of the right-propagating (or faster) wave—namely, the 2-DSW—possibly emanating from the dam break in Eqs.~\eqref{eq:riemann}. In the discussion below, we assume a positive dispersion curvature (which fixes the orientation and polarity of the 2-DSW to $-1$ each, respectively; see for details \cite{el_dispersive_2016-1}), but the discussions can be adapted suitably for a negative dispersion curvature. Furthermore, we assume that a point of zero-density (i.e., the emergence  of cavitation points) is not attained within the 2-DSW interior. 

The underlying strategy for the characterization of the 2-DSW is based on the 
so-called DSW-fitting method \cite{el_resolution_2005,hoefer_shock_2014}, which provides insight into the solitonic and linear edge speeds while circumventing the need to determine the full integral curve of the two-phase Whitham modulation equations~\cite{whitham2011linear}. The details relating to applying the procedure to the DNLS can be found in our earlier work \cite{mohapatra2025dam}. We state instead, only the relevant equations to compute the edge speeds of the 2-DSW. 

{We then can find the following zero-amplitude integral curve ODE describing the ``dynamics" of the linear edge (two-phase) wavenumber $\overline{\kappa}$
\begin{equation}\label{eq: Integral curve for k}
    \frac{d\overline{\kappa}}{d\overline{k}} = \frac{\partial_{\overline{k}}\omega_0^{+}}{A\left(\overline{k}\right) - \partial_{\overline{\kappa}}\omega_0^{+}},
\end{equation}
where $A(\overline{k})=$\\$-2\hat{\omega}_1L\left(\sin\left(\overline{k}L\right) + \left(E\left(\frac{\overline{k}L}{2},2\right)+\frac{\sqrt{w_{11}}}{\sqrt{-2\hat{\omega}_1}}\left|c_+\right|\right)\sqrt{\cos\left(\overline{k}L\right)}\right)$ is the second, one-phase Whitham velocity ($\lambda^{(2)}$) with the 2-DSW jump condition incorporated. Furthermore, the ODE in Eq.~\eqref{eq: Integral curve for k} is subject to  the ``initial" condition $\overline{\kappa}(k_0) = 0$ (i.e., the wavenumber prescription at the solitonic edge \cite{el_resolution_2005}). }
Note that the dispersion relation whose partial derivatives appear in Eq.~(\ref{eq: ODE at solitonic edge})
is given above in Eq.~(\ref{eq: linear dispersion relation}).
Furthermore, $k_0$ and $\rho_0$ are the intermediate background Stokes wavenumber and density respectively, that satisfy {Eqs.~\eqref{density} and \eqref{velocity}}. 
Having ascertained $\kappa_{+}=\overline{\kappa}(k=0)$, we can then compute the group velocity (and linear edge speed) as 
\begin{equation}\label{eq: group velocity}
    s^{+} = \partial_{\overline{\kappa}}\omega_0^{+}\left(\kappa^{+},k=0\right).
\end{equation}
The solitonic edge fitting can be carried out in a similar manner, but by first introducing the conjugate “wavenumber” $\tilde{\kappa}$ and “angular frequency” $\tilde{\omega}_0^{\pm}(\tilde{\kappa})=-i\omega_0^{\pm}(i\tilde{\kappa})$, which offer a more convenient framework than the wave amplitude, particularly since no universal amplitude modulation equation exists \cite{el_resolution_2005}. 
The (zero-wavenumber) integral curve ODE at the solitonic edge reads
\begin{equation}\label{eq: ODE at solitonic edge}
    \frac{d\tilde{\kappa}}{d\overline{{k}}} = \frac{\partial_{\overline{k}}\tilde{\omega}_0^{+}}{A\left(\overline{k}\right) - \partial_{\tilde{\kappa}}\tilde{\omega}_0^{+}}.
\end{equation}
{Specifying the initial condition $\tilde{\kappa}(0)=0$ (i.e. the zero amplitude prescription at the linear edge) to the ODE Eq.~\eqref{eq: ODE at solitonic edge}, we obtain the conjugate wavenumber at the solitonic edge ($\tilde{\kappa}_{-}=\tilde{\kappa}(k_0)$).} In turn, this yields a quasi-analytical expression for the soliton edge speed
\begin{equation}
    s^{-}=\frac{\tilde{\omega}_0(\tilde{\kappa}_-,k_0)}{\tilde{\kappa}}.
\end{equation}
\subsection{Weakly nonlinear 2-DSW}
\label{DNLS-WNL}
To explore how weak yet finite disturbances propagate over uniform backgrounds, we examine asymptotic reductions that emerge when the leading nonlinear effects are quadratic and counterbalance long-wave ($\kappa \ll1$) dispersion. This is precisely the regime where the KdV framework and its higher-order weakly nonlinear variants naturally appear. {Extending these reductions beyond the classical KdV level ({whose DSWs were first studied in \cite{gurevich1974nonstationary}}) can shed light on the other novel shock waveforms that can emerge.} 

In this vein, we have two relevant quasi-continuum reductions to the tight-binding approximation. The first amongst them is given by the (right-traveling) KdV reduction
\begin{equation}\label{eq: KdV reduction for tight-binding model}
   a^{(1)}_{\tau} +\Tilde{\beta}a^{(1)}a^{(1)}_{\xi} - \tilde{\alpha}_3a^{(1)}_{\xi\xi\xi} = 0,
\end{equation}
where $\Psi\sim |c|+\epsilon a^{(1)}(\xi,\tau)$ ($\epsilon \ll |c|$) represent the weak modulations on a homogeneous background of amplitude $|c|$. Furthermore, $\xi=\epsilon^{1/2}(x-\tilde{\alpha}_1t), \hspace{3mm} \tau = \epsilon^{3/2}t$. This choice of scaling asserts a balance between the quadratic nonlinearity and third order dispersion.
Moreover, the quantities $\tilde{\alpha}_j$ represent the coefficients arising in the long-wave asymptotic expansion of the linear dispersion relation associated with two-phase waves, i.e.,
\begin{align}
    &\omega_0^{+}(\kappa;k,\rho_0,\hat{\omega}_1,L)\sim \tilde{\alpha}_1(k,\rho_0,\hat{\omega}_1,L)\kappa +\\\nonumber&\tilde{\alpha}_3(k,\rho_0,\hat{\omega}_1,L)\kappa^3+\tilde{\alpha}_5(k,\rho_0,\hat{\omega}_1,L)\kappa^5+\cdots
\end{align}
Here, $\tilde{\alpha}_1=\lambda_2$ (the dispersionless speed), while the higher-order terms characterize the dispersive corrections at various orders. Notably, the phase (and group) velocities of propagating disturbances decrease as $\hat{\omega}_1$ reduces (i.e., as the potential depth parameter $V_0$ increases). Consequently, as $V_0$ increases, the likelihood of site-to-site tunneling diminishes, which is an anticipated outcome. Moreover, $\tilde{\beta}(\rho_0,k,L,\hat{\omega}_1)$ is the coefficient of quadratic nonlinearity. Finally, we mention that the sign of $\tilde{\alpha}_3$ determines the polarity and orientation of a generated DSW. We refer the reader to \cite{kamchatnov2004dissipationless} for the details on the derivation.

\begin{figure*}
    \centering
    \includegraphics[width=\linewidth]{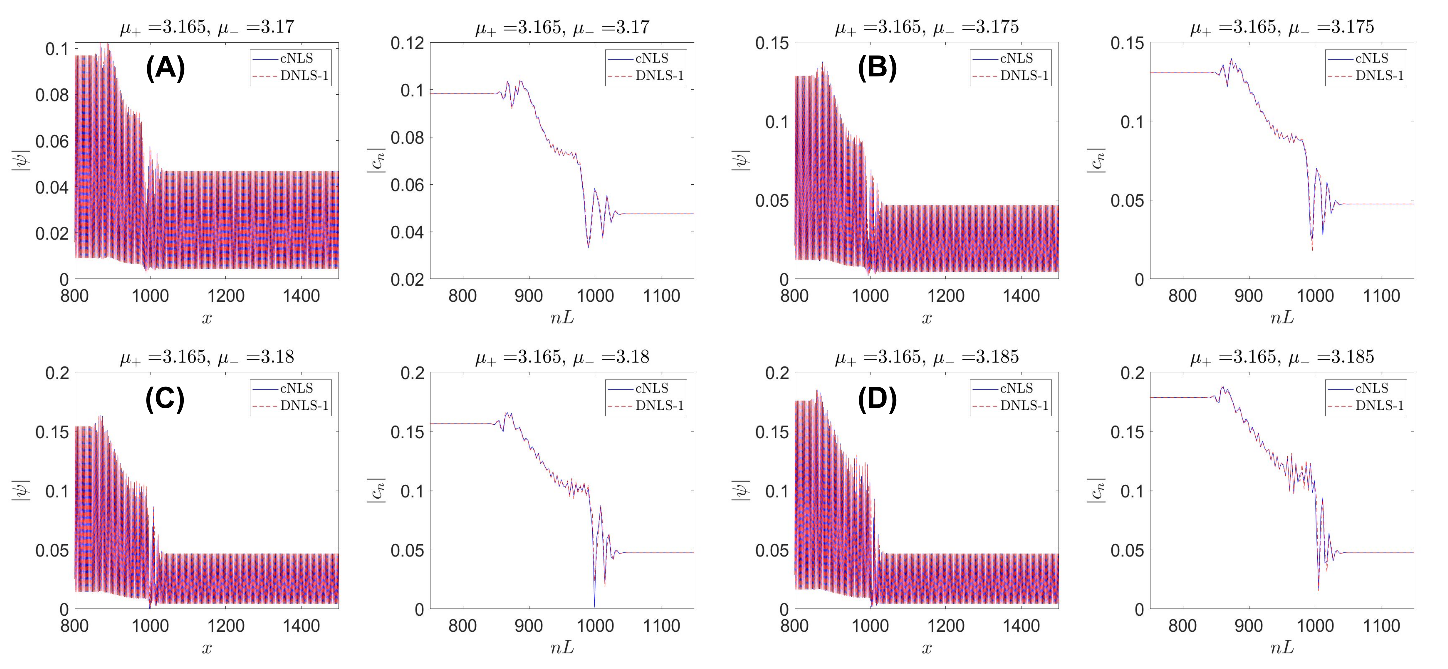}
    \caption{A catalog of dam-break results at $t=1000$, for fixed $\mu_{+}$, while varying $\mu_{-}<3.19$, illustrating the generation of a radiating, right-propagating DSW for $V_0=12$. The dam break at $t=0$ (location of the initial step) occurs at $x_n=nL=301\pi$. Within each of the 4 panels, there are sub-panels showing the dynamics at the continuum level (left) and discrete level, (i.e. at the locations coinciding with the potential minima) (right). The results from the continuum NLS simulations in both sub-panels (Eq.~\eqref{eq: Periodic dNLS}) have been shown in blue solid line, with its tight-binding approximation (DNLS, Eq.~\eqref{system-wanner}) in red dashed line. As $\mu_{-}$ increases, there is an eventual disagreement, particularly at the solitonic edge (B,C), which approaches the vacuum point ($|\psi|=0$). Moreover, in panel (D), while the NLS and its tight-binding approximation show reasonable agreement, the DSW exhibits prominent leftward radiation, possibly due to significant higher-order dispersive effects.}
    \label{fig:1}
\end{figure*}
In order to characterize the (right propagating) shock waveforms in the limit of small jumps, we consider the Riemann problems posed to Eq.~\eqref{eq: KdV reduction for tight-binding model} 
\begin{equation}
 a^{(1)}(\xi,0)=\begin{cases}
0,;\xi<0\\
-1,;\xi>0.
\end{cases}
\end{equation}
with the coefficients of the quasi-continuum reductions being defined on the intermediate hydrodynamic background $(\sqrt{\rho_0},k_0)$. The shock waves emanating from these step-like initial data can then provide an approximation to those from the tight-binding approximation (after performing the relevant spatio-temporal rescaling, and assuming $\sqrt{\rho_0}\sim \mathcal{O}(1)$) (see, e.g., \cite{mohapatra2025dam,hoefer_shock_2014}).
\section{Wave patterns emerging from the Generalized Riemann initial data}
\label{wavepatt}
In this section, we study the family of generalized dam breaks (Eq.~\eqref{eq: Initial condition for inhomogeneous NLS}) of the continuum NLS (cNLS) in Eq.~\eqref{eq: Periodic dNLS}. As shown in Fig.~\ref{fig:periodic_construction}, each nonlinear periodic state $u^{\mp}(x)$ is an elementary superposition of Wannier states (see also Eq.~\eqref{eq: Initial condition for inhomogeneous NLS}). 

We check the fidelity of the tight-binding approximation in capturing the dynamics of the continuum NLS at later times. To do so, we map the problem to the dual Wannier ``space". {At the outset, it is important to reiterate the multi-step reduction of Eq.~\eqref{system-wannercomplex} to the tight-binding reduction, involving the approximation of the discrete convolution (linear term) and the nonlinear term therein \cite{Alfimov_2002,kevrekidis2009discrete}. To begin, restricting attention to the first band ($\alpha=1$) and focusing on the discrete convolution term, we note that the relative strength of next-nearest-neighbor interactions compared to the nearest-neighbor ones is governed by the ratio of the Fourier cosine coefficients $\hat{\omega}_2/\hat{\omega}_1$. This ratio is beyond all algebraic orders small in $V_0^{-1}$ (with $\hat{\omega}_0 \gg \hat{\omega}_1 \gg \hat{\omega}_2 \cdots$ for $V_0^{-1} \ll 1$), which justifies neglecting all but the nearest-neighbor contributions in the convolution in the regime of $V_0$'s considered herein. In the vein of the nonlinear terms in Eq.~\eqref{system-wannercomplex}, we consider the ratio of the overlapping matrix elements embodying the strength of nonlinear interactions across sites $w^{n n_1 n_2 n_3}_{1 \alpha_1\alpha_2\alpha_3}/w^{n n n n}_{1 \alpha_1\alpha_2\alpha_3}$, which is exponentially small in $V_0^{-1}\ll 1$. A further approximation—namely, the omission of terms with $w^{nnnn}_{1 \alpha_1 \alpha_2 \alpha_3}$ for which $\alpha_2 \neq 1$ and $\alpha_3 \neq \alpha_1$—is justified by noting that the corresponding contributions are rapidly oscillatory and thus have a negligible effect on the time-averaged dynamics for $t \sim V_0$. Thus, the overlapping matrix elements of leading order importance over long times include $w^{nnnn}_{1\alpha_1 1\alpha_1}$. Furthermore, there is an additional decay in $V_0^{-1}$ of the ratio $w^{nnnn}_{1\alpha_1 1\alpha_1}/w^{nnnn}_{1111}$ (strength of $\alpha_1$-$1$ interband interactions), thus leading to the scalar, tight-binding reduction in Eq.~\eqref{system-wanner}. The ability to control the errors at each approximation level lends strong support, for the values of $V_0 \gg 1$ that we consider, to the tight-binding approximation as a faithful first model of the continuum NLS dynamics subject to a periodic potential.} Some quantitative details regarding the
comparisons of the terms that we neglect with respect to those that we
retain can be found in~\cite{Alfimov_2002}. Here, we will also justify these
assumptions a posteriori through the detailed comparison of the cNLS and
the DNLS results.

At $t=0$, {the mapping to Wannier space} yields the Riemann initial condition (or dam break) in Eq.~\eqref{eq:riemann} to the tight binding approximation. The system of ODEs at the lattice sites $x_n=nL$ as given by the Wannier reduction is then evolved to a final time $t$ to yield $c_n(t)$. The approximation to the cNLS dynamics at $t$ can then be reconstructed by taking the superposition of the time-evolved Wannier basis coefficients
\begin{equation}
    \label{eq:Wannier_approx_cont_dyna}
    \psi(x,t)\approx \sum_{n} c_n(t) w_n(x).
\end{equation}

A comparison between the cNLS dynamics and its tight-binding DNLS-based counterpart [through Eq.~\eqref{eq:Wannier_approx_cont_dyna}] is referred to as the continuum-level comparison.

Alternatively, to test the accuracy of the tight-binding approximation, we could construct a single-band Wannier function decomposition of the time-evolved continuum cNLS dynamics, $\psi(x,t)$, i.e., determine the exact $c_n^{\rm exact}(t)$ by projecting $\psi(x,t)$ onto the 
eigenmodes $w_n(x)$.
We refer to this as being a comparison made at the {\it discrete level} between the two models. At the outset, we mention two sources of discrepancy between the cNLS and the tight-binding approximation dynamics
\begin{itemize}
    \item Discrepancies between the one-phase dispersion function $\Omega(k;L)$ and the band structure $\nu_{1}(k)$ can be mitigated by incorporating higher-order Fourier coefficients $\hat{\omega}_j$ ($j>1$) into the tight-binding  approximation. 
    In such cases, extending the approximation to include interactions with next-nearest neighbours (at sites $n \pm 2$) often yields a more accurate description and we refer to as the “second” tight-binding approximation, whose one-phase dispersion relation is given by $\Omega^{(2)}(k;L)=\hat{\omega}_0+2\hat{\omega}_1\cos(kL)+2\hat{\omega}_2\cos(2kL)$. 
    \item The fundamental assumption of ignoring the contributions to the dynamics from the higher-energy bands ($\alpha>1$). These contributions are typically minimized for very deep lattices $V_0\gg1$, which will be the focus of our work. In future work, it would also be interesting to explore a reduced-order vector tight-binding approximation that incorporates intraband energy transfer \cite{Alfimov_2002}. Indeed, this would lead to a multi-component
    DNLS lattice that would be of interest to explore in its own right.
\end{itemize}
To fix ideas, we display comparisons of generalized dam breaks for $V_0=12$. Here, the Fourier cosine coefficients assume the values $\hat{\omega}_0\approx 3.19$, $\hat{\omega}_1\approx -0.0123$ and $\hat{\omega}_2\approx 8.66\times10^{-5}$, indicating the accuracy of a nearest-neighbor tight-binding reduction in capturing $\nu_1(k)$. In this regime, the first-band Wannier functions are also nearly Gaussian in shape (Fig.~\ref{fig:periodic_construction}(B)). 

A palette of results (at $t=1000$) for this potential depth is shown in Fig.~\ref{fig:1}. The right background $u^{+}(x)$ is fixed, with its associated eigenvalue close to the band edge $\nu_1(0)$, while the left periodic background is varied (by tuning the eigenvalue $\mu_-$). The left subpanel in each of the four panels compares the dynamics of the tight-binding approximation (DNLS) with those of the cNLS at the continuum level. As mentioned before, this continuum waveform is reconstructed by a weighted superposition of Wannier functions (with $c_n(t)$ being the corresponding weights obtained from the time evolution of the tight-binding approximation). On the other hand, the right subpanel in each of the four panels compares the cNLS and DNLS dynamics at the discrete level. For each of the dam breaks, we observe good agreement throughout the entire evolution in the counterpropagating rarefaction–radiating DSW waveforms, with the best correspondence obtained for the smallest $\mu_-$. As $\mu_-$ increases, the DSW emits stronger radiation, a feature attributable to higher-order dispersive effects \cite{sprenger2017shock,mohapatra2025dam}.

\begin{figure}
    \centering
    \includegraphics[width=\linewidth]{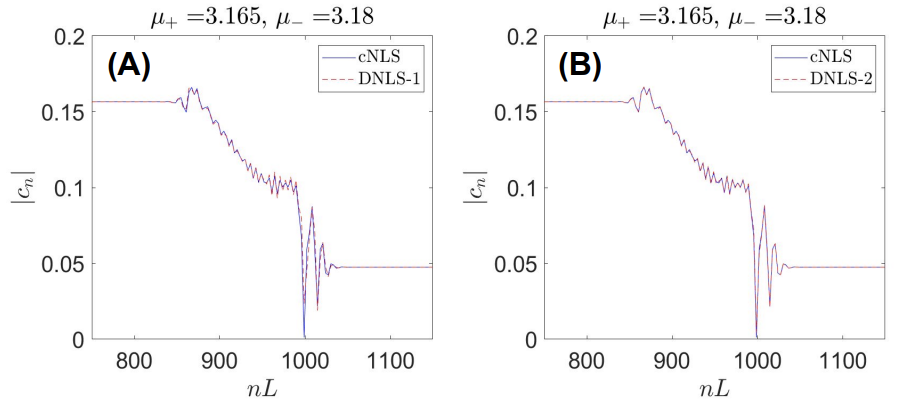}
    \caption{{A comparison of the results at $t=1000$ is presented at the discrete level, showing the continuum NLS dynamics (blue solid line) alongside the discrete tight-binding model (red solid line) with nearest-neighbor interactions (A) (Eq.~\eqref{system-wanner-full}), and the corresponding comparison with the tight-binding model including next-nearest-neighbor interactions (B) (Eq.~\eqref{system-wanner}, with $n_1=2$). The tight-binding model with next-nearest-neighbor interactions exhibits higher fidelity to the continuum predictions throughout the entire evolution, \textit{especially} at the solitonic edge of the DSW.}}
    \label{fig:2}
\end{figure}
\begin{figure}
    \centering
    \includegraphics[width=\linewidth]{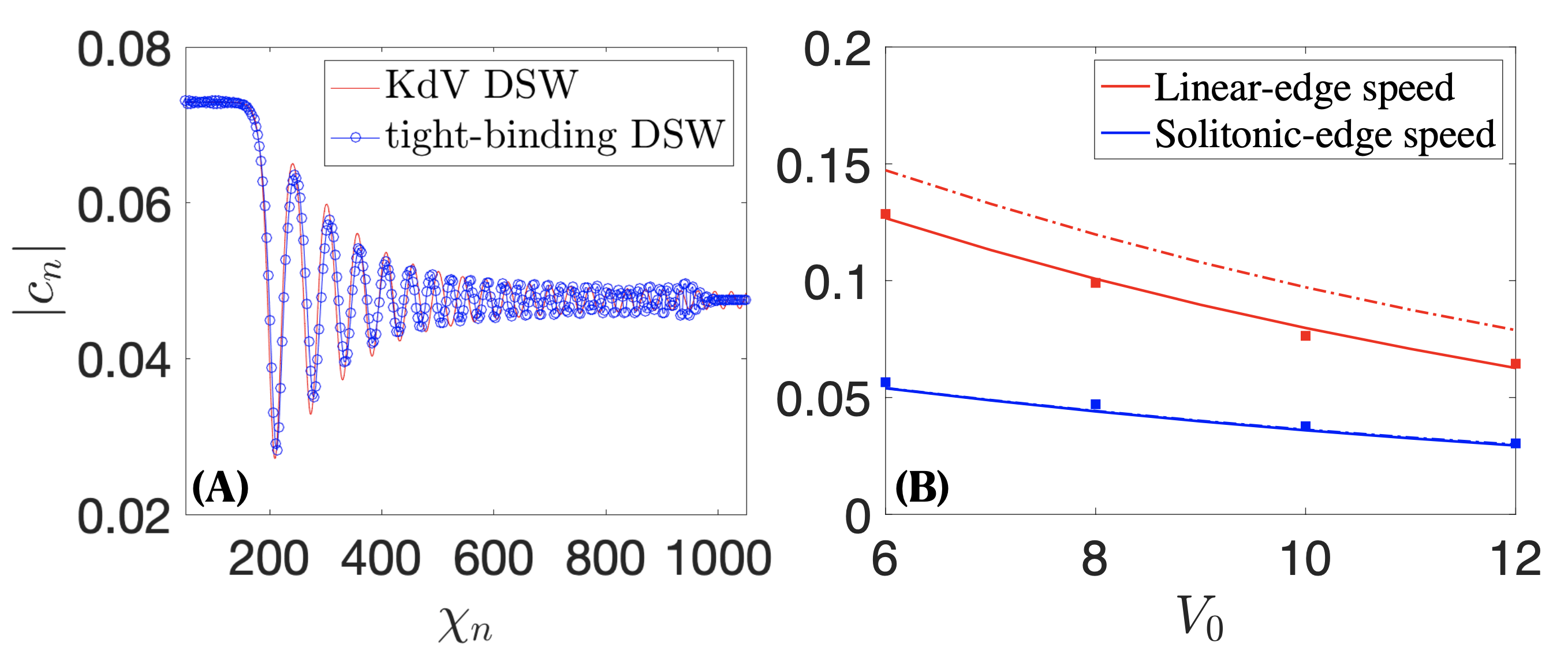}
    \caption{(A) Comparison of the asymptotic KdV DSW  as reconstructed
    at the level of the DNLS via Eq.~(\ref{eq: KdV reduction for tight-binding model}) 
    (and discussion therearound) with a small-amplitude tight-binding DSW generated using $c_-\approx 0.1$, $c_+\approx 0.05$, and $V_0=6$, showing good agreement. $\chi_n=x_n-n_0 L$ is an appropriate shifted coordinate, where $n_0L$ is the location of the initial step. 
(B) DSW edge speeds as a function of the potential depth $V_0$, with $c_-\approx 0.1$ and $c_+\approx 0.05$ held fixed. Squares represent data extracted from numerical simulations, solid curves show the quasi-analytical results obtained from the DSW fitting procedure, and dashed curves indicate the asymptotic KdV DSW predictions (Eqs.~\eqref{edge-speeds-DSW-KDV}). {Note that the solitonic-edge speed predicted by the KdV reduction (dashed blue in the right panel) coincides with the DSW fitting result (solid blue line in the right panel).} 
}
    \label{fig:KdV}
\end{figure}

\begin{figure*}
    \centering
    \includegraphics[width=\linewidth]{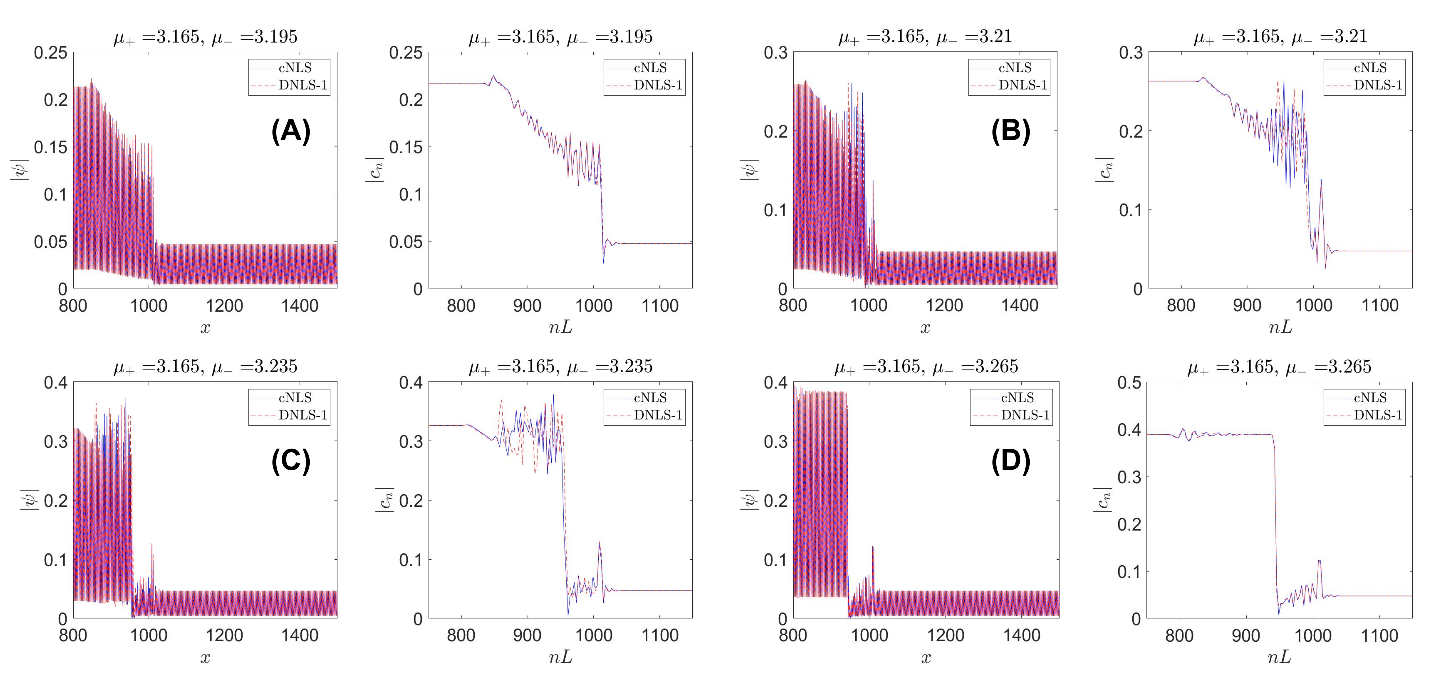}
    \caption{A palette of dam-break results at $t=1000$, for fixed $\mu_{+}$, while varying $\mu_{-}> 3.19$ (with $V_0=12$), illustrating the generation of structures quite distinct from the right-propagating DSW, shown in Fig.~\ref{fig:1}. The dam break at $t=0$ (location of the initial step) occurs at $x_n=nL=301\pi$. Within each of the 4 panels, there are sub-panels showing the dynamics at the continuum level (left) and discrete level, (i.e., at the locations coinciding with the potential minima) (right). The results from the continuum NLS simulations in both sub-panels (Eq.~\eqref{eq: Periodic dNLS}) have been shown in blue solid line, with its tight-binding approximation (DNLS, Eq.~\eqref{system-wanner}) in red dashed line. With increasing $\mu_-$, the right propagating traveling DSW in (A) eventually breaks down, due to a generalized two-phase modulational instability (MI) (see (B)). For even larger parameter values in $\mu_-$, the interaction of the high-genus structure generated by modulational instability begins to annihilate the left-propagating rarefaction flow through mutual interaction (C). Finally for sufficiently large $\mu_-$, a heteroclinic, (stationary) breathing feature is generated.}
    \label{fig:3}
\end{figure*}
In order to understand the source of discrepancy between the tight-binding approximation (Eq.~\eqref{system-wanner}) and the cNLS dynamics for increasing $\mu_-$, we invoke a tight-binding approximation involving next-nearest neighbors interaction (DNLS-2), i.e., $n-2\leq n_1\leq n+
2$ in Eq.~\eqref{system-wanner-full}. Clearly, this approximation provides a more accurate representation of the first-band dispersion relation and, consequently, is expected to exhibit higher fidelity to the cNLS. To illustrate this, we present a discrete-level comparison (at $t=1000$) for the parameter set $\mu_-=3.18$, $\mu_+=3.165$ in Fig.~\ref{fig:2}. The DNLS-2 results are seen to be virtually indistinguishable from those of the cNLS, not only at $t=1000$ (as shown) but throughout the entire waveform's evolution.
The success of the tight-binding approximation in particular is noteworthy, considering that the simulation time for the cNLS dynamics on a local workstation is $\sim 10000$ s, while that for the tight-binding approximation (DNLS or DNLS-2) is $\sim 100$ s.

On a related note, we also highlight the quasi-continuum reductions of the tight-binding approximation (see Sec.~\ref{DNLS-WNL}), which may offer an analytical framework for investigating the coherent structures that emerge during the evolution of the underlying Riemann problem. 
In this vein, a universal model for small-amplitude DSWs is the KdV equation (Eq.~\eqref{eq: KdV reduction for tight-binding model}), which arises from an asymptotic balance between weak third-order dispersion and quadratic nonlinearity. For small-amplitude jumps, we compare the KdV DSW with the DNLS (and, subsequently, the cNLS) modulations, observing good agreement. To illustrate this, we explore comparisons made within the parametric window of $6\leq V_0\leq 12$ for the right-propagating (and small amplitude) DSW. The results of such comparisons are shown in Fig.~\ref{fig:KdV}, where we display a family of Riemann problems corresponding to $c_-\approx 0.1$ and $c_+\approx 0.05$ (a small-amplitude jump), for varying potential depths $6\leq V_0\leq 12$. At the outset, we remark that within this parametric window, the DSWs obtained from the cNLS and DNLS models are in good agreement. In Fig.~\ref{fig:KdV}(A), the corresponding DSW generated by the KdV, as reconstructed via the multiple scale reduction 
of Sec.~\ref{DNLS-WNL} (see, in particular Eq.~(\ref{eq: KdV reduction for tight-binding model})
and the discussion around it),
is compared with that of the DNLS at $t=3000$, showing similarly good agreement. This, in turn, highlights the utility of the KdV framework for analyzing modulations of periodic waves in the cNLS setting. While the solitonic edge is captured quite accurately, a slight discrepancy is observed near the linear edge. In Fig.~\ref{fig:KdV}(B), the edge speeds of the tight-binding reduction (depicted using squares) are compared to the quasi-analytical predictions from the DSW fitting procedure, as shown in the solid curves (Eqs.~\eqref{eq: Integral curve for k} and \eqref{eq: ODE at solitonic edge}) and the KdV predictions (dashed curves). {The corresponding KdV predictions for the solitonic ($v_-$) and linear edge speed ($v_+$) can be shown to be related to the jump across the DSW, $\Delta =\sqrt{\rho_0}-|c^{+}|$, where $\sqrt{\rho_0}=(|c^{-}|+|c^{+}|)/2$ is the amplitude of the intermediate hydrodynamic background. Through an appropriate rescaling of the well-known formula for the (non-dimensionalized) KdV-DSW edge speeds  (see also \cite{el_resolution_2005}), the associated edge speeds for Eq.~\eqref{eq: KdV reduction for tight-binding model} are
\begin{equation}
\label{edge-speeds-DSW-KDV}
    v_-=-\frac{2\tilde{\beta}}{3}\Delta+\tilde{\alpha}_1,\;v_+=\tilde{\beta}\Delta+\tilde{\alpha}_1.
\end{equation} }

Aside from a slight discrepancy (within 15 $\%$ across the entire parameter range) in the linear-edge speeds, the KdV prediction provides an a very good estimate of the small-amplitude, propagating DSW in the periodic lattice.

We now turn to the right-propagating wave patterns that arise for sufficiently large jumps $|\mu_- - \mu_+|$. In this vein, we fix $V_0=12$ and $\mu_+=3.165$, and explore the parametric window $\mu_->3.19$ (see Fig.~\ref{fig:3}). At the outset, we note slight quantitative discrepancies between the DNLS (and DNLS-2) and the cNLS dynamics. These slight deviations may stem from coupling between energy bands, an effect not captured by the standard tight-binding approximation. A natural extension for future work would be to investigate vector DNLS systems that incorporate interband coupling through cross-phase modulation terms \cite{Alfimov_2002}. However, we emphasize that this minor quantitative discrepancy does not detract from the overall qualitative agreement, particularly regarding the delineation of distinct phenomenological regimes, which we describe below.

{Within the parametric window considered, we no longer observe a radiating DSW, but instead, a traveling DSW (tDSW) \cite{sprenger2017shock} structure (Fig.~\ref{fig:3}(A)). For this parameter set, the coefficients of third- and fifth-order dispersion in the long-wave expansion are found to be $\tilde{\alpha}_3(\rho_0,k_0)\approx 0.0048$ and $\tilde{\alpha}_5(\rho_0,k_0)\approx -0.185$, respectively. This indicates that higher-order dispersive effects are expected to play a significant role. Moreover, the amplitude scale (defined by the ratio of the step to the amplitude of the intermediate background $|\sqrt{\rho_0}-|c^{+}||/\sqrt{\rho_0}$ ) here is $\epsilon\approx 0.5$, where besides the higher-order dispersion, higher-order nonlinear effects are also expected to play a role, which can possibly be explained through an extended KdV model \cite{baqer2025shallow}. We leave such characterizations for future considerations.} The tDSW waveforms, for larger $\mu_-$, break down over relatively shorter evolutionary times, possibly due to \textit{generalized} modulational instability of two-phase wavetrains (Fig.~\ref{fig:3}(B)). {A recent work (c.f. \cite{sprenger2023whithammodulationtheorytwophase}) has systematically investigated the notion of (in)stability of two-phase periodic waves in a family of nonlinear Schrödinger-type models (of which the DNLS is a member) with dispersion governed by pseudodifferential operator(s) (i.e., full-dispersion models). These two-phase periodic waves correspond to homogeneous solutions of the so-called two-phase Whitham modulation equations. As such, their instability can then be viewed in the framework of the Whitham modulation equations, wherein it reduces to a modulational instability problem. 
The characterization of this generalized two-phase MI for the DNLS has also played a role in recent studies of DSW dynamics \cite{panayotaros2016shelf,mohapatra2025dam}. } 
{The nonlinear stage of this two-phase MI gives rise to an oscillatory regime featuring a high-genus structure (see also Ref.~\cite{mohapatra2025dam}), the detailed understanding of which merits further investigation in future work.} The interaction of this oscillatory feature with the left-propagating rarefaction wave is eventually seen to result in the annihilation of the latter (see Fig.~\ref{fig:3}(C)). For even larger values of $\mu_{-}$, the critical jump required for the existence of rarefaction waves is exceeded (see Eq.~\eqref{Intermediate-wave-dens}), and a stationary yet breathing heteroclinic structure emerges. The pulsations of this feature, in turn, drive small-amplitude, right-propagating excitations (see Fig.~\ref{fig:3}(D) and Ref.~\cite{mohapatra2025dam}).


\section{Conclusions and Scope for future work}
\label{conclusions}
In the present work, we investigated the formation of dispersive shock waves (DSWs) and rarefaction waves in periodic lattices governed by Eq.~\eqref{eq: Periodic dNLS}. Specifically, these waveforms are interpreted as long-time asymptotic states arising from a one-parameter family of step-like initial data that connect two distinct periodic lattice eigenmodes, with the right eigenmode held fixed. The selected eigenmodes bifurcate from a band edge with positive dispersion curvature, thereby ensuring their modulational stability.

To clarify the connection with DSWs in homogeneous media, we invoke a tight-binding approximation that reduces the continuum NLS to a discrete tight-binding nonlinear Schrödinger (DNLS) model in the deep-lattice limit, where the gaps between consecutive energy (or dispersion) bands are sufficiently large. Dispersive shock waves have been studied extensively in discrete models \cite{mohapatra2025dam,konotop1997dark,panayotaros2016shelf,kamchatnov2004dissipationless}, and our work therefore establishes an explicit connection to this body of literature within the broader context of non-convex, discrete dispersive hydrodynamics.

This connection, in turn, reveals a variety of additional excitations that emerge in the regime where standard DSWs cease to exist, generating, as a result, structures such as
traveling DSWs and heteroclinic breathing waveforms, among others. Further effects, such as the breakdown of tDSW structures due to the generalized modulational instability of two-phase periodic waves, are also captured within this framework (see \cite{mohapatra2025dam} for related examples of such effects and \cite{sprenger2023whithammodulationtheorytwophase} for the conceptual framework of two-phase MI).

Quantitatively, we show that the hierarchy of tight-binding models, \textit{progressively} succeeds in capturing phenomena that depend sensitively on the first-band dispersion relation, including DSW cavitation (where the solitonic edge reaches zero amplitude), the parametric regimes supporting propagating tDSWs, and their eventual breakdown, among others. The slight quantitative discrepancy observed in the extreme regime of modulational instability of two-phase periodic waves can be attributed to energy transfer to higher dispersion bands, motivating the exploration of a relevant vector DNLS  system as suggested in \cite{Alfimov_2002}. Such a vector DNLS could then be employed to study periodic lattice DSWs in the intermediate $V_0$ regime, providing a comprehensive yet computationally efficient approach—typical simulation times are roughly one-hundredth of those required for the corresponding continuum simulations—. This has the
potential to enable investigating these excitations across the range from intermediate to large $V_0$. 
It does not escape us that these gains in decreased simulation times
may be accordingly more significant in higher-dimensional settings.
However, we expect this framework of vector DNLS models to provide a qualitative (and less accurate quantitatively) description in the shallow periodic potential limit.

Finally, we stress the experimental relevance and applicability of our approach, both in the context of optical systems, including waveguide arrays and photorefractive crystal lattices \cite{jia2007dispersive,Lederer2008_PhysRep_DiscreteSolitons}, as well as in Bose-Einstein condensates confined within optical lattices \cite{brazhnyi2004theory,morsch}, where
recently significant advances have been made into the visualization of localized
states~\cite{cruickshank2025single}. 
Due to the rapidly advancing study of non-convex dispersive hydrodynamics, we anticipate that the coherent shock structures revealed in our numerical investigations could not only be observed experimentally but also understood 
in further quantitative detail. Relevant studies are presently in progress and will be
reported in future publications.

{\bf Acknowledgements.}
This research was supported by the U.S. National Science Foundation under the awards DMS-2204702 and PHY-2408988 (PGK). This research was partly conducted while P.G.K. was 
visiting the Okinawa Institute of Science and
Technology (OIST) through the Theoretical Sciences Visiting Program (TSVP). 
This work was also 
supported by a grant from the Simons Foundation
[SFI-MPS-SFM-00011048, P.G.K]. 

\bibliographystyle{unsrt}
\bibliography{main}
\end{document}